# Highly accurate fiber transfer delay measurement with large dynamic range

**J. W. Dong,**[1,2] **B. Wang,**[1,2,*] **C. Gao,**[1,2] **Y. C. Guo,**[1,2] **and L. J. Wang**[1,2,3]

[1]*Joint Institute for Measurement Science, Tsinghua University, Beijing 100084, China*
[2]*State Key Laboratory of Precision Measurement Technology and Instruments, Department of Precision Instruments, Tsinghua University, Beijing 100084, China*
[3]*Department of Physics, Tsinghua University, Beijing 100084, China*
[*]*bo.wang@tsinghua.edu.cn*

**Abstract:** A novel and efficient method for fiber transfer delay measurement is demonstrated. Fiber transfer delay measurement in time domain is converted into the frequency measurement of the modulation signal in frequency domain, accompany with a coarse and easy ambiguity resolving process. This method achieves a sub-picosecond resolution, with an accuracy of 1 picosecond, and a large dynamic range up to 50 km as well as no measurement dead zone.

**References and links**

1. B. Mukherjee, "WDM optical communication networks: progress and challenges," IEEE J. Sel. Areas Comm. **18**(10), 1810–1824 (2000).
2. B. Lee, "Review of the present status of optical fiber sensors," Opt. Fiber Technol. **9**(2), 57–79 (2003).
3. K. Predehl, G. Grosche, S. M. F. Raupach, S. Droste, O. Terra, J. Alnis, Th. Legero, T. W. Hänsch, Th. Udem, R. Holzwarth, and H. Schnatz, "A 920-kilometer optical fiber link for frequency metrology at the 19th decimal place," Science **336**(6080), 441–444 (2012).
4. W. Shillue, "Fiber distribution of local oscillator for Atacama Large Millimeter Array," presented at OFC/NFOEC, San Diego, CA USA, 24–28 Feb. 2008.
5. B. Wang, X. Zhu, C. Gao, Y. Bai, J. W. Dong, and L. J. Wang, "Square kilometer array telescope - precision reference frequency synchronisation via 1f-2f dissemination," Sci. Rep. **5**, 13851 (2015).
6. https://www.ptb.de/emrp/neatft_publications.html.
7. W. Ng, A. A. Walston, G. L. Tangonan, J. J. Lee, I. L. Newberg, and N. Bernstein, "The first demonstration of an optically steered microwave phased array antenna using true-time delay," J. Lightwave Technol. **9**(9), 1124–1131 (1991).
8. B. Vidal, T. Mengual, C. Ibáñez-López, J. Martí, I. McKenzie, E. Vez, J. Santamará, F. Dalmases, and L. Jofre, "Simplified WDM optical beamforming network for large antenna arrays," IEEE Photon. Technol. Lett. **18**(10), 1200–1202 (2006).
9. D. Dolfi, F. Michel-Gabriel, S. Bann, and J. P. Huignard, "Two-dimensional optical architecture for phase and time-delay beam forming in a phased array antenna," Opt. Lett. **16**(4), 255–257 (1991).
10. B. Vidal, T. Mengual, and J. Marti, "Optical beamforming network based on fiber-optical delay lines and spatial light modulators for large antenna arrays," IEEE Photon. Technol. Lett. **18**(24), 2590–2592 (2006).
11. S. R. Jefferts, M. A. Weiss, J. Levine, S. Dilla, E. W. Bell, and T. E. Parker, "Two-way time and frequency transfer using optical fibers," IEEE Trans. Instrum. Meas. **46**(2), 209–211 (1997).
12. C. Lopes, and B. Riondet, "Ultra precise time dissemination system," *Frequency and Time Forum, 1999 and the IEEE International Frequency Control Symposium, 1999., Proceedings of the 1999 Joint Meeting of the European* (IEEE, 1999), pp. 296–299.
13. M. Rost, M. Fujieda, and D. Piester, "Time transfer through optical fibers (TTTOF): progress on calibrated clock comparisons," in *Proceedings of 24th European Frequency and Time Forum*, paper 6.4 (2010).
14. B. Wang, C. Gao, W. L. Chen, J. Miao, X. Zhu, Y. Bai, J. W. Zhang, Y. Y. Feng, T. C. Li, and L. J. Wang, "Precise and continuous time and frequency synchronisation at the 5×10-19 accuracy level," Sci. Rep. **2**, 556 (2012).
15. J. Kalisz, "Review of methods for time interval measurements with picosecond resolution," Metrologia **41**(1), 17–32 (2004).
16. D. L. Philen, I. A. White, J. F. Kuhl, and S. C. Mettler, "Single-mode fiber OTDR: experiment and theory," IEEE Trans. Microwave Theory Tech. **30**(10), 1487–1496 (1982).
17. Q. Bing, T. Andrew, Q. Li, and L. Hoi-Kwong, "High-resolution, large dynamic range fiber length measurement based on a frequency-shifted asymmetric Sagnac interferometer," Opt. Lett. **30**(24), 3287–3289 (2005).

18. L. D. Nguyen, B. Journet, I. Ledoux-Rak, J. Zyss, L. Nam, and V. V. Luc, "Opto-electronic oscillator: applications to sensors," in *Proceedings of IEEE International Meeting on Microwave Photonics/2008 Asia-Pacific Microwave Photonics Conference* (IEEE, 2008), pp. 131–134.
19. K. H. Yoon, J. W. Song, and H. D. Kim, "Fiber length measurement technique employing self-seeding laser oscillation of fabry–perot laser diode," Jpn. J. Appl. Phys. **46**(1), 415–416 (2007).
20. I. Fujima, S. Iwasaki, and K. Seta, "High-resolution distance meter using optical intensity modulation at 28 GHz," Meas. Sci. Technol. **9**(7), 1049–1052 (1998).
21. Y. L. Hu, L. Zhan, Z. X. Zhang, S. Y. Luo, and Y. X. Xia, "High-resolution measurement of fiber length by using a mode-locked fiber laser configuration," Opt. Lett. **32**(12), 1605–1607 (2007).
22. K. Yun, J. Li, G. X. Zhang, L. L. Chen, W. J. Yang, and Z. G. Zhang, "Simple and highly accurate technique for time delay measurement in optical fibers by free-running laser configuration," Opt. Lett. **33**(15), 1732–1734 (2008).

## 1. Introduction

Due to its low attenuation, high reliability and accessibility, optical fiber has become an attractive transmission medium for different application areas, such as optical communication [1], fiber-optic sensing [2], and many large-scale scientific or engineering facilities [3–6]. Fiber transmission induced time delay measurement is indispensable in these applications. For the application of phased array antenna, fiber transfer delay (FTD) measurement accuracy directly affects its control accuracy, and further affects the beam forming results [7–10]. For time and frequency synchronization network, FTD measurement is a key step to realize clock synchronization [11–14]. Ordinarily, the time delay is measured in time domain, such as using the time interval counter (TIC) [15] or optical time domain reflectometer (OTDR) [16], but they have drawbacks of low accuracy and existing dead zones. Recently, efforts have been made to meet requirements of large dynamic range and high accuracy based on propagation delay measurement in frequency domain [17–20]. The common method is converting the FTD measurement into the longitudinal mode spacing measurement of a mode-locked fiber laser or the beating frequency measurement of a ring cavity. A mode-locked fiber laser based FTD measurement method was demonstrated. It achieved a resolution of a few centimeters with a dynamic range of hundreds of kilometers [21]. In another report, a technique that included the fiber under test (FUT) as a part of a ring-cavity fiber laser and measured the high-order harmonic beating frequency without mode locking, offers an accuracy of $10^{-8}$ for a fiber length of 100km and of $10^{-6}$ for a several-meters-long one [22]. Although there is no need for mode locking or laser stabilization, the difficulty of its ambiguity resolving process increases with the FUT length growth, which makes the measurement time-consuming and inconvenient.

In this Letter, we propose and demonstrate a novel and simple technique for FTD measurement by transferring a microwave signal modulated laser light. In this way, the FTD measurement is converted into the precision measurement of microwave signal's frequency, accompany with a coarse and easy ambiguity resolving process. Compared with previous methods, this is far more convenient and the ambiguity resolving procedure no longer depends on the fiber length. When the microwave signal is frequency-locked to the transfer delay, the long-term and real-time FTD measurement can be implemented through continuous frequency measurement. We demonstrate an FTD measurement system with a large dynamic range up to 50 km as well as no measurement dead zone. For long-term FTD measurement of a 2 m long fiber, we obtained a resolution of 0.2 ps and an accuracy of 1 ps. For short-term FTD measurement of a 50 km long fiber, we obtained a resolution better than 0.3 ps.

## 2. Methods

Fig. 1 shows the schematic of the FTD measurement system. During measurement, one end of the FUT is connected to a Faraday mirror, and the other end is connected to the FTD measurement system through a fiber circulator. In this way, the FUT is included as a part of the FTD measurement loop, and the FUT will be double-passed by a probe light. Here, the probe light is a microwave signal modulated 1550 nm laser light. The microwave signal is provided by an oscillator, which consists of a 100MHz voltage control oscillator (VCO) and a frequency multiplier (×12). Through mixing and filtering operations on the microwave signals before and after transmission, an error signal $V_e \propto \cos(\phi_p)$ is obtained. Here $\phi_p$ denotes the phase delay induced by transmission and can be expressed as

$$\phi_p = 2\pi f \cdot t = 2\pi f \cdot \frac{L_{op}}{c}, \qquad (1)$$

where $f$ is frequency of the microwave signal, $t$ is the transfer delay, $L_{op}$ is the optical path length of the FTD measurement loop, and $c$ is the velocity of light in vacuum. Through a phase locked loop, the frequency of the microwave signal $f$ can be locked onto the transfer delay $t$, making $V_e \propto \cos(\phi_p) = 0$. Thus $\phi_p$ is a constant as

$$\phi_p = \left(N + \frac{1}{2}\right)\pi. \qquad (2)$$

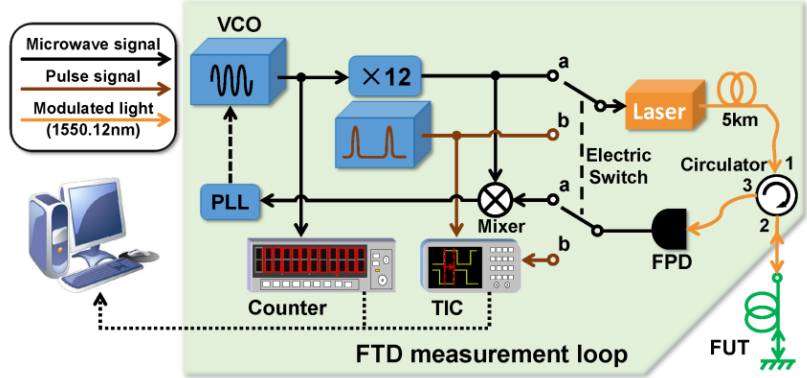

Fig. 1. Schematic of FTD measurement system. VCO, 100 MHz voltage control oscillator; FPD, fast photo detector; FUT, fiber under test; PLL, phase locked loop; TIC, time interval counter. The microwave signal and the pulse signal are switched to modulate the amplitude of a 1550 nm laser light. After transferred in the 5 km long internal fiber, the laser light is coupled into the FUT through an optical circulator. The reflected light is detected by a FPD. The pulse signals before and after transmission are used for coarse FTD measurement by a TIC. The microwave signals before and after transmission are used to generator an error signal through frequency filtering and mixing operations. The PLL uses the error signal to control the frequency of the VCO, making the microwave signal frequency-locked to the transfer delay.

Here, $N$ is an integer and can be considered as an ambiguity of the measurement. According to Eq. (1) and (2), we can get

$$t = \frac{2N+1}{4f}. \qquad (3)$$

The transfer delay $t$ can be obtained as long as $f$ and $N$ are determined. The frequency $f$ can be precisely measured by a frequency counter. In order to get $N$, a general time delay discrimination method is used for FTD coarse measurement. The same 1550 nm laser light is modulated by a pulse signal transferring through the FTD measurement loop. To rapidly switch between precise and coarse measurement, a double-pole-double-throw electric switch is inserted into the FTD measurement loop (shown in Fig. 1). Pulse signals before and after transmission are used to measure the coarse transfer delay $t_{coarse}$ by a TIC. Based on Eq. (3), $N$ can be obtained by equation

$$N = 2ft_{coarse} - \frac{1}{2}. \tag{4}$$

To evaluate the ambiguity resolution of the method, according to Eq. (4), we can analyze the uncertainty of $N$

$$\Delta N = 2f \cdot \Delta t + 2t \cdot \Delta f. \tag{5}$$

Here, $f$ is 1.2 GHz in our measurement system. $\Delta f$ is decided by the frequency counter accuracy, which is better than $10^{-10}$ for a commercial meter. $\Delta t$ is determined by coarse FTD measurement accuracy. For common measurement, the FUT length is below 100 km and the transfer delay $t$ will not exceed 1 ms. Consequently, the second term on the right-hand side of Eq. (5) is small enough to be neglected, and the uncertainty of $N$ is mainly limited by the coarse FTD measurement accuracy which deteriorates with the FUT length growth. In order to obtain sufficient enough accuracy of $N$, i.e. $\Delta N \leq 0.5$, $\Delta t$ should be less than 210 ps, which is easy to realize using commercial TIC.

Once $N$ is determined, $t$ can be obtained according to Eq. (3). Here, the measured FTD $t$ contains not only double-passed FUT delay, but also the propagation delays of electronic circuits, cables, fiber pigtails of the measurement system, which belong to system delay ($t_0$). Consequently, to get one-way FUT delay ($t_F$), we need to calibrate the system delay. For convenience of system delay measurement and eliminating the dead zone, an internal fiber (~5 km long) is preset into the FTD measurement system. When no FUT is connected, the modulated laser light entering into the port 1 of fiber circulator is reflected by the end face of the port 2 (FC/PC connector) and transferred back to measurement system. In this way, the system delay $t_0$ is measured and the one-way FUT delay $t_F$ can be obtained by following relation

$$t_F = \frac{1}{2}(t - t_0). \tag{6}$$

For the case of long-term continuous measurement, temperature variations and mechanical vibrations may cause fluctuation of the system delay. Consequently, a system delay control (SDC) loop is used to monitor and compensate the system delay fluctuation, making it stable during the measurement. The FTD measurement loop and the SDC loop are combined together as shown in Fig. 2. Another 1547 nm laser light is modulated by a 1 GHz signal referenced by Hydrogen-Maser. The modulated 1547 nm laser light is coupled into the internal preset fiber and transfers together with the 1550 nm laser light. After transmission, two laser lights are separated by a wavelength division multiplexer (WDM). The 1547 nm laser light is detected by a fast photo detector (FPD2). Via frequency mixing and filtering operations on the 1 GHz signals before and after transmission, an error signal proportional to the system phase delay is obtained. A part of the internal fiber (~2 km long) is wrapped around a copper wheel. A PI controller uses the error signal to cancel out the variation of the system delay by changing the spool's temperature with a sensitivity of 87 ps/℃ and a total dynamic range of 1.7 ns. In this way, the fluctuation of the system delay is well compensated.

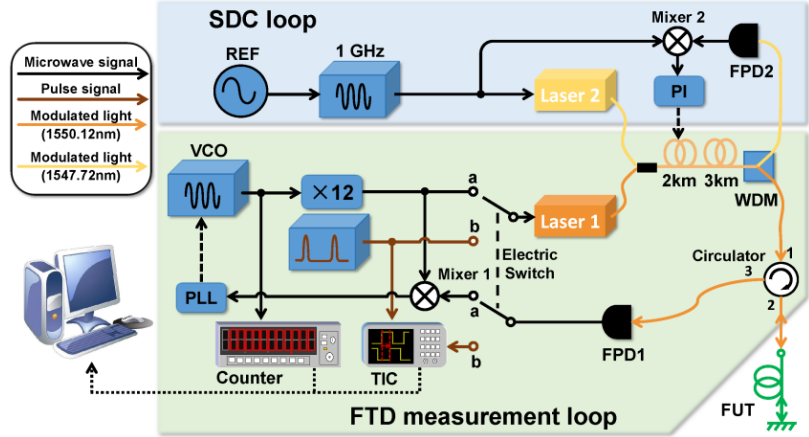

Fig. 2. Schematic of FTD measurement system with SDC loop. WDM, 1547/1550 nm wavelength division multiplexer; REF, reference signal; PI, proportional integral controller. A 1547 nm laser light modulated by a 1 GHz signal is coupled into the internal preset fiber together with the 1550 nm laser light through a 50/50 optical coupler. Two laser lights are separated by a WDM. The 1547 nm laser light is detected by FPD2. Via frequency mixing and filtering operations, an error signal proportional to the system phase delay is obtained. Through changing the temperature of a part of the internal fiber (~2 km long), a PI controller uses the error signal to cancel out the variation of the system delay.

During measurement, when the SDC loop is locked, the system delay is stabilized. When the FTD measurement loop is locked, $N$ is fixed as a constant. In this way, a long-term, real-time FTD measurement can be implemented by continuous measurement of frequency.

## 3. Results and discussion

To evaluate the system delay stability, we performed a series of system delay measurements without connecting FUT. Fig. 3 shows the measured system delay fluctuations and the converted system stability via time deviation (TDEV). The black line is the result when the system internal fiber is running freely, and the red line is the result when the system delay fluctuation is compensated. We observed the fluctuation of $\pm 100\,ps$ for the uncompensated loop. While, for the compensated loop, it is reduced to $\pm 1\,ps$, shown in Fig. 3(a). A significant improvement can be clearly observed in the corresponding TDEV plot, shown in Fig. 3(b). The values of compensated loop are always below 210 fs, whereas in freely running system, the TDEV reaches 30 ps for averaging times of $10^3$ s. It indicates that with system delay compensation, the long-term stability is kept at sub-picosecond level. As a comparison, we used TIC to measure the stabilized system delay by transferring a pulse signal. The result is shown in Fig. 4 (black line). A delay fluctuation of $\pm 25 ps$ can be seen. Comparing measurement results of two methods, we note that the uncertainty of the measurement using the present method is reduced by more than one order of magnitude compared to that of using TIC.

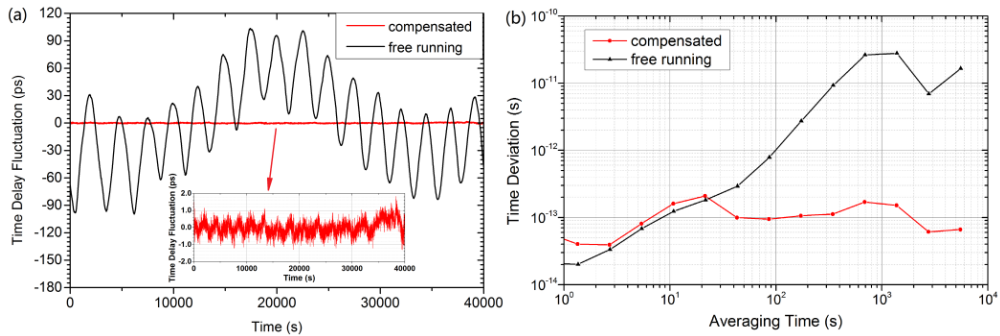

Fig. 3. Measurement results of the uncompensated and compensated system delays. (a) The system delay fluctuation. The black line is the result when the system internal fiber is running freely, showing a fluctuation of $\pm 100\,ps$. The red line is the result with the system delay fluctuation compensated, showing a fluctuation of $\pm 1\,ps$. (b) The time deviation of the system delay derived from the measured system delay fluctuation.

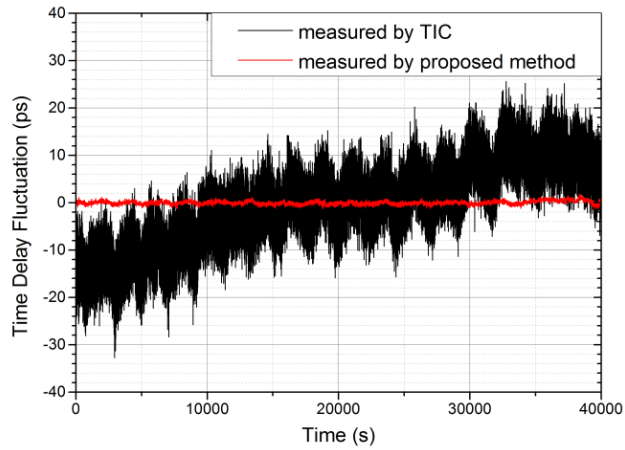

Fig. 4. Measurement results of the compensated system delay fluctuations using two methods. The black line is the result measured by TIC, showing a fluctuation of $\pm 25 ps$. The red line is the result measured by proposed method, showing a fluctuation of $\pm 1 ps$.

To further verify the accuracy of the FTD measurement system, we used it to measure the FTD of a 2 m long fiber which can be considered as a constant. We repeated tests in different times of a day. The result is shown in Fig. 5. It can be seen that, the means of all measured

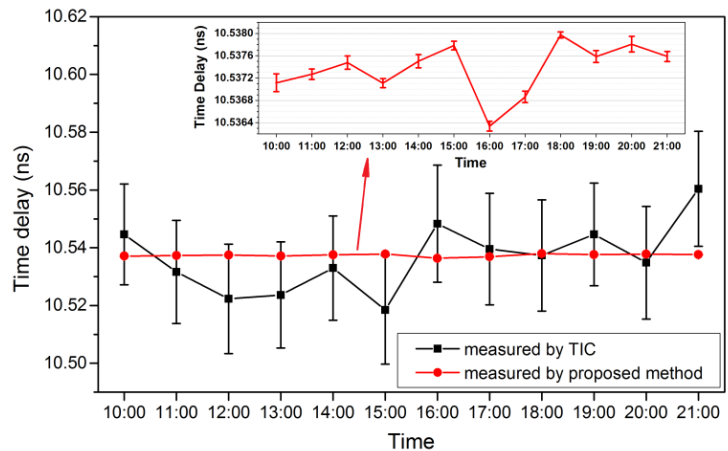

Fig. 5. FTD measurement results of a 2 m long fiber. The averaging period is 100 s. The error bar is the standard deviation of each measurement. The black line is the FTD measured by TIC, showing a long-term fluctuation of $\pm 20 ps$ with an uncertainty of 20 ps. The red line is the FTD measured by proposed method. The long-term fluctuation is reduced to $\pm 1 ps$ with an uncertainty below 0.2 ps.

FTDs using two methods are well overlapped. The statistical error of the proposed method is below 0.2 ps and the long-term fluctuation of it is below $\pm 1 ps$, whereas in the measurement using commercial TIC, the uncertainty reaches approximately 20 ps and the long-term fluctuation increases to $\pm 20\,ps$. It indicates that there are two orders of magnitude improvement in the measurement resolution, and the measurement accuracy is obviously improved as well.

To demonstrate the measurement range, we also measured fibers with different lengths. For a fiber about 50 km long, we obtained the time delay of $25192235.2 \pm 0.2\,ps$. In particular, a delay variation of less than 0.3 ps can be precisely monitored caused by the temperature fluctuation while taking measurement. This result indicates that the proposed method has a large dynamic range at least up to 50 km, with an extremely high resolution.

**4. Conclusion**

We have demonstrated a novel and efficient scheme for fiber transfer delay measurement. Using this method, continuously real-time FTD measurement can be realized. More importantly, the method has a sub-picosecond high resolution, with a large dynamic range of more than 50 km, and a high accuracy, as well as no dead zone. Owing to its good repeatability, high reliability, and simple measurement process, the method may find wide applications, particularly in large-scale, phased-array antenna such as the Square Kilometre Array (SKA) for radio astronomy.

**Acknowledgments**

We acknowledge financial support from the National Key Scientific Instrument and Equipment Development Project (No. 2013YQ09094303) and the Beijing Higher Education Young Elite Teacher Project (No. YETP0088).